\documentclass[fleqn,usenatbib]{mnras}

\usepackage[T1]{fontenc}

\DeclareRobustCommand{\VAN}[3]{#2}
\let\VANthebibliography\thebibliography
\def\thebibliography{\DeclareRobustCommand{\VAN}[3]{##3}\VANthebibliography}

\usepackage{graphicx}	
\usepackage{amsmath}	
\usepackage{amssymb}	
\usepackage{newtxtext,newtxmath}
\usepackage{subfig}
\usepackage{multirow}
\usepackage{xcolor}

\newcommand{\nustar}{\textit{NuSTAR}}
\newcommand{\obj}{Swift~J0243.6+6124}

\newcommand{\ergps}{erg\,s$^{-1}$}
\newcommand{\ergpspcm}{erg \,s$^{-1}$\,cm$^{-2}$}

\title[\obj~ observations with \nustar]{ULX pulsar \obj\ observations with  \nustar -- dominance of reflected emission in the super-Eddington state}

\author[Bykov et al.]{
S.D. Bykov$^{1,2}$\thanks{E-mail:sergei.d.bykov@gmail.com}, M.R. Gilfanov$^{1,3}$, S.S. Tsygankov$^{4,3}$ and  E.V. Filippova$^{3}$
\\
$^{1}$Max Planck Institute for Astrophysics, Karl-Schwarzschild-Str 1, Garching b. München D-85741, Germany\\
$^{2}$Kazan Federal University, 18 Kremlyovskaya street, Kazan, Russia\\
$^{3}$Space Research Institute, Russian Academy of Sciences, Profsoyuznaya 84/32, 117997 Moscow, Russia\\
$^{4}$Department of Physics and Astronomy, FI-20014 University of Turku,  Finland}

\date{Accepted XXX. Received YYY; in original form ZZZ}

\pubyear{2022}

\begin{document}
\label{firstpage}
\pagerange{\pageref{firstpage}--\pageref{lastpage}}
\maketitle

\begin{abstract}

We report the discovery of the bright reflected emission component in the super-Eddington state  of the ULX pulsar \obj{}, based on the \nustar{} observations of the source during its 2017 outburst. The flux of the reflected emission is  weakly variable over the pulsar phase while the direct emission shows significantly larger pulsation amplitude. 
We propose that in this system the neutron star finds itself in the centre of the well formed by the inner edge of the geometrically thick super-Eddington accretion disc truncated by the magnetic field of the pulsar. The aspect ratio of the well is $H/R\sim 1$. The inner edge of the truncated disc is continuously illuminated by the emission of the accretion column giving rise to the  weakly variable reflected emission. As the neutron star rotates, its emission sweeps through the line of sight, giving rise to the pulsating direct emission. From Doppler broadening of the iron line, we measure the truncation radius of the accretion disc $\sim~50~R_{\rm g}$. The inferred dipole component of the magnetic field  is consistent with previous estimates favouring a not very strong  field. The uniqueness of this system is determined by its moderately super-Eddington accretion rate and the moderate magnetic field so that the inner edge of the truncated geometrically thick accretion disc is seen from the neutron star at a large solid angle.

\end{abstract}




\section{Introduction}
\label{intro}
The X-ray transient \obj~ was discovered  by \textit{Swift} observatory during its   2017 outburst \citep{Cenko2017}. Pulsations with a $9.8$~sec period were detected \citep{Kennea2017}, establishing the source as an accreting neutron star and a transient X-ray pulsar (XRP).  
The outburst lasted  $\sim200$~ days, after which the source transitioned to the regime of flaring activity. 
The optical counterpart was found to be a Be star \citep{Kouroubatzakis2017, Reig2020} confirming that the source belongs to the class of  Be X-ray binaries (BeXRB). From \textit{Gaia} data, the distance to the source was estimated to be $6.8^{+1.5}_{-1.1}$~ kpc  \citep{Bailer-Jones2018}; the mean value of the distance value is adopted below. 
At the peak of the outburst, the source reached the luminosity of  $\sim~2~\cdot~ 10^{39}$~\ergps{} (bolometric) which  exceeds the Eddington limit  for a neutron star by a factor of $\sim~10$ \citep{Tsygankov2018, Doroshenko2020}, thus placing the source in a cohort of Ultra-Luminous X-ray  pulsars (ULX pulsars). Remarkably, it is  the first of this kind found in our galaxy.

Due to its uniqueness, X-ray studies of \obj\    are plentiful.  
\citet{Wilson-Hodge2018} investigated the spectral and timing behaviour of the  pulsar with \textit{Fermi}/GBM and \textit{NICER} telescopes. 
They found that the pulse profile shows strong evolution and energy dependence, with a transition from single- to double-peaked shape (and vice-versa) as the pulsar brightens (fades). Such behaviour is typical for transient X-ray pulsars and has been predicted theoretically \citep{Basko1975}. From the luminosity at which  the pulse profile change happens they estimated the magnetic field of the neutron star to be $B>10^{13}$~G. However, based on the absence of the propeller effect  at the declining phase of the outburst, \citet{Tsygankov2018, Doroshenko2020} argued that the magnetic field should be smaller.  \citet{Kong2022} reported the discovery of the variable cyclotron scattering absorption feature (CSRF) in the spectrum of this pulsar (in the energy range 120--140 keV). The implied magnetic field strengths is therefore $\sim1.6\times10^{13}$ G, and it is believed to originate from the non-dipole component of the magnetic field.

\textit{Insight-HXMT} provided  broadband X-ray data with high cadence \citep{Zhang2019, Kong2020, Wang2020}. 
Using these data \citet{Doroshenko2020} reported specific changes in timing and spectral properties of the pulsar.   These changes were interpreted as an onset of super-Eddington accretion via the gas pressure dominated disc and the following transition to the radiation pressure supported disc (RPD). These results are important since they show that not only do radiation patterns change with pulsar luminosity, but also surrounding matter distribution may evolve in the course of the outburst. The presence of the RPD indicates that the magnetic field may not be very large. This claim is furtherly substantiated by the presence of pulsations and variability of X-ray flux in a quiescent state observed by \nustar{} \citep{Doroshenko2020}.

The iron line complex was studied extensively by \citet{Jaisawal2019} using \textit{NICER} and \nustar\  data. They detected several bright lines of  asymmetric shape in the $6-7$~keV range. In particular, data could be fitted with the model consisting of lines at $6.4$, $6.7$ and $6.98$~keV from neutral and ionized iron species. In their model, the $6.4$~keV line was very broad ($\sigma$ up to $1$~keV). They interpreted these results that the iron emission is produced in the accretion disc or plasma trapper in the magnetosphere. 
\citet{Tao2019} also reported the presence of  the significant reflection component and relativistic line shapes  in \obj.

The peak luminosity of \obj{} lies  between 'normal' transient XRPs and bright ULX pulsars. Significant changes in the matter distribution around a magnetized neutron star were predicted to take place at  high mass accretion rates  \citep{Mushtukov2017}, and \obj\  may be in the regime where such a transition may happen.

Notwithstanding ample observational data, only one phase-resolved spectroscopic study was done with \nustar{} data in the sub-Eddington state of the pulsar \citep{Jaisawal2018}. This paper aims to fill this gap and is dedicated to the phase-resolved spectroscopy of \nustar{} data  in the high luminosity state, with particular emphasis on the behaviour of the iron line complex. In section~\ref{nu_data} we describe the data reduction and analysis.  In sect.~\ref{results-timing} we discuss the timing analysis, in~\ref{results-ph-ave} we describe the results of phase-averaged spectra, and in~\ref{results-ph-res} we report on the phase-resolved spectroscopy. We discuss our results in sect.~\ref{disussion} and conclude in sect.~\ref{conclusion}.

\section{\nustar\, data}
\label{nu_data}
\nustar\, \citep{Harrison2013} is a grazing incidence hard X-ray telescope sensitive in the 3--79~keV energy range with the energy resolution of 4\% at 10~keV. It consists of two identical modules: FPMA and FPMB, data from each being processed separately. In the course of its 2017-2018  type II outburst, \obj\, was observed with \nustar\,  five times between MJD~58000 and MJD~58200, observations  9030231900(2,4,6,8) and 90401308001.  Observation 90401308001 was excluded due to its short exposure.  We add observation 90401308002 of one of the subsequent mini-outburst (likely Type I outburst) to our data set. We use the following aliases for observations: 90302319002 -- I, 90302319004 -- II, 90302319006 -- III, 90302319008 -- IV, 90401308002 -- V. Dates of the observations are the following: I - MJD~58032; II - MJD~58057; III - MJD~58067; IV - MJD~58094; V - MJD~58369.

The \nustar\,  data reduction was carried out using the {\sc HEASOFT} v. 6.28 package and  \nustar\, CALDB files (v117). Standard data-processing tools were used to extract level 2 ({\sc nupipeline}) and 3 ({\sc nuproducts}) products. We choose source and background regions for data extraction: the former is centred on the source position, while the latter is placed as far from the source as possible. Both regions have a circular shape with a 180~arcsec radius. Because of the high count rate in all observations we set a bitmask "STATUS==b0000xxx00xxxx000" for {\sc nupipeline} as recommended\footnote{\url{https://heasarc.gsfc.nasa.gov/docs/nustar/analysis/}}. All data were barycentred,  but the time of arrival of photons is not corrected for the orbital motion of the system due to the lack of orbital solution based on all spin frequency data of the source. Since the on-source time of \nustar\ observations is relatively short  (a few tens of ks) compared to the orbital period of $\sim 30$~days this should not present any errors in the timing analysis. Spectra from both modules were simultaneously fit between 4 and 79~keV with {\sc xspec} package \citep{Arnaud1996}. For phase-resolved spectroscopy photons were folded into 10 phase bins. All \nustar\, spectra were grouped  to have at least 25~counts per energy bin and $\chi^2$ statistics was used. The error intervals are given for $90\%$ confidence.

\begin{figure}
    \centering
    \includegraphics[width=0.5\textwidth]{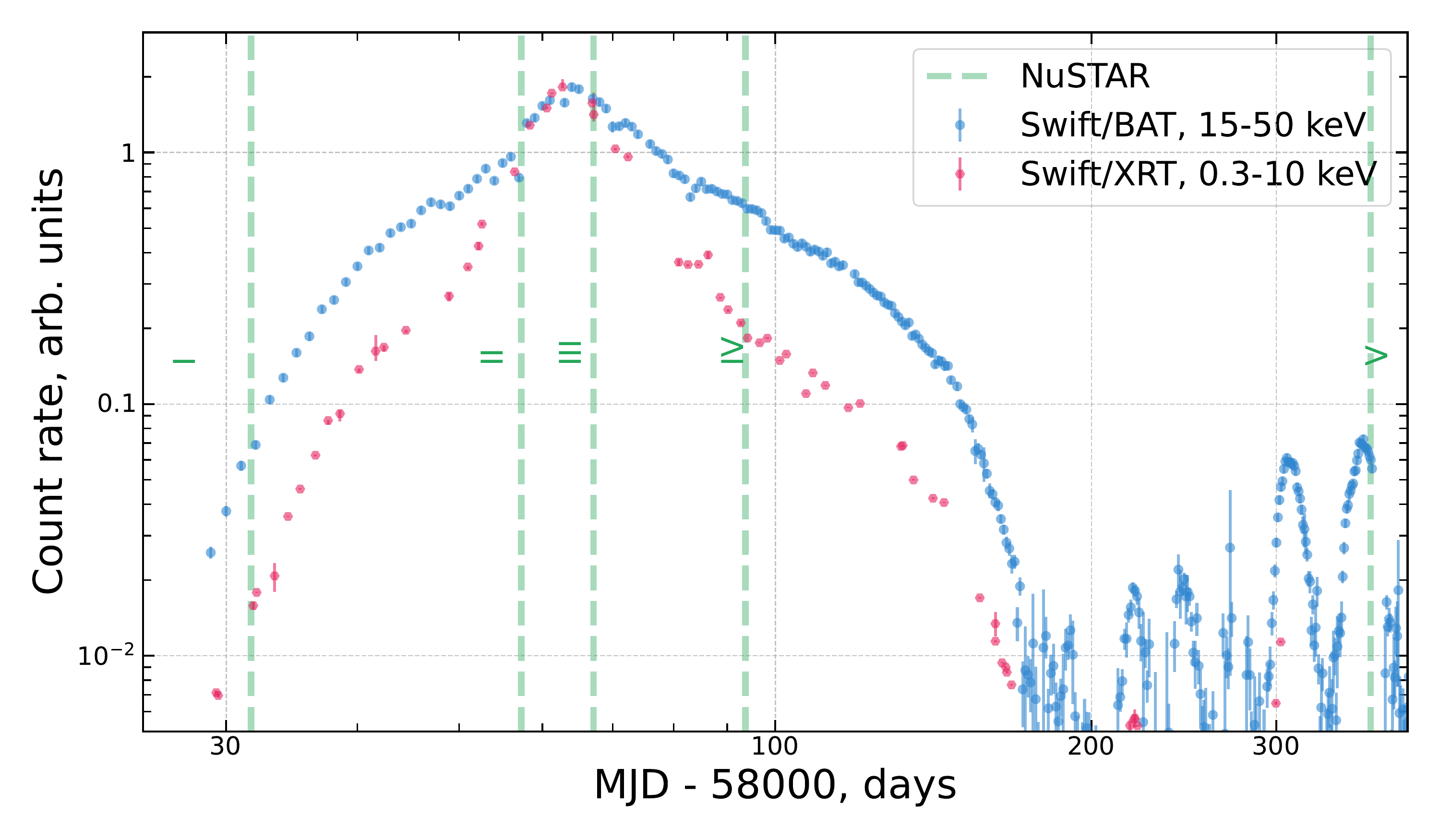}
    \caption{\textit{Swift}/BAT and XRT light curves of a source in arbitrary units of count rate (blue and green symbols respectively). \nustar\ observations are marked with green vertical lines. Note the time axis in the log scale in units of days after MJD 58000.}
    \label{fig:lightcurve}
\end{figure}

\section{Results}

Fig.~\ref{fig:lightcurve} shows the long term light curve of the source in the 15--50~keV band (\textit{Swift}/BAT data\footnote{\url{https://swift.gsfc.nasa.gov/results/transients/weak/SwiftJ0243.6p6124/}}) and the 0.3--10~keV range (\textit{Swift}/XRT data\footnote{reduced from an online tool  \url{https://www.swift.ac.uk/user_objects/}, see \citep{Evans2007}}). 
The main, giant, outburst lasts approx 200~days and is followed by a series of smaller flares (possibly Type I outbursts). Dates of \nustar{} observations are marked with vertical lines. All observations except one are taken during the main outburst (I--IV, from left to right), and one during the smaller flare (V, the right-most one in the figure). The source was in a sub-Eddington state in observations I and V, and  in  a super-Eddington state in observations II, III and IV (see, e.g.,  fig. 1 in \citet{Wang2020}).

\subsection{Timing analysis}
\label{results-timing}

Pulsations of X-ray flux are detected in all observations. For each observation, we determine the period of the pulsar  with {\sc efsearch} routine, using combined  data of both \nustar{} detectors. The  periods are  9.85425, 9.84435, 9.82340, 9.80105, and 9.7918~seconds for observations  I, II, III, IV and V respectively (Solar system barycenter). These values of periods were used for folding the data for pulse profiles and phase-resolved spectroscopy. Due to the lack of a reliable orbital solution for the binary system no phase connection was made to synchronize pulse profiles between different observations. 

\begin{figure}
    \centering
    \includegraphics[width=0.5\textwidth]{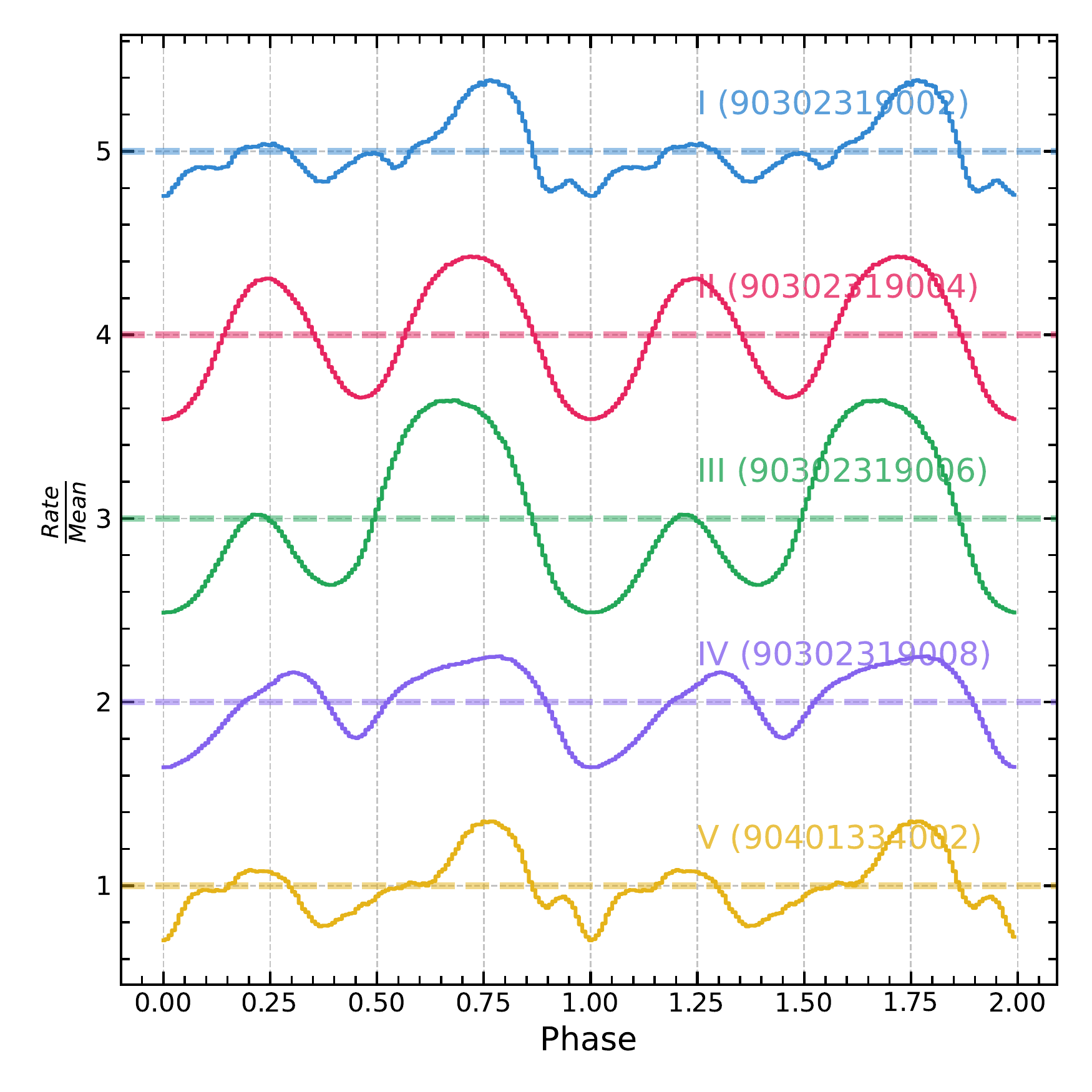}
    \caption{Pulse profiles (normalized to mean count rate) in the 4--79~keV energy range. The profiles were offset from each other by $1.0$ for readability.  The time since the start of the outburst goes from the top to the bottom. All pulse profiles except for the first and the last were taken in the super-Eddington state of the pulsar. Profiles were shifted so that the main minimum is at phase 0.  }
    \label{fig:pulse-profiles}
\end{figure}

In Fig.~\ref{fig:pulse-profiles} we show the pulse profiles in the 4--79~keV energy range of all five \nustar{} observations. Pulse profiles are normalized to the mean count rate (per observation). At the beginning of the outburst (observation I), the pulse profile has a complex shape with multiple minor peaks, but overall the profile is dominated by the one-peak component.  The last observation (V) taken during another, minor flare, but in similar brightness,  shows a very similar shape.  The pulsed fraction PF \footnote{defined as $PF = \frac{\operatorname{max}C - \operatorname{min}C}{\operatorname{max}C +  \operatorname{min}C}$ of the pulse profile $C$} is $\sim30\%$ in the considered energy range.

As the source becomes more luminous, the pulse profile changes drastically. In the two  observations where the source was the brightest (II and III), the pulse profiles show a very smooth and symmetric double-peaked shape. In   observation II the peaks are roughly equal in amplitude and the pulsed fraction is $\sim45\%$, whilst in  observation III one peak  dominates the overall shape and the pulsed fraction is somewhat larger, $\sim55\%$. In the next observation, IV, the pulse profile features two asymmetric peaks of roughly equal height. The pulsed fraction is $\sim30\%$.  The dependence of PF on the energy of photons may be found in \citet{Tao2019} or \citet{Wang2020}. Note that as we did not do phase-connection between individual observations, Fig.~\ref{fig:pulse-profiles} may not correctly represent how \textit{a particular pulse profile maximum} evolves in time, but rather shows  the general evolution of the shape of the pulse profile  \citep[see][]{Doroshenko2020,Jiren2022}.

\subsection{Phase-average spectral analysis}
\label{results-ph-ave}
The X-ray spectra of \obj\ measured by various observatories have been extensively analysed in the works cited in section~\ref{intro} and the main  parameters for common pulsar spectral models  have been obtained and discussed. Below we focus on several facts and features, which were not considered in the previous analyses.

\begin{figure}
    \centering
    \includegraphics[width=0.5\textwidth]{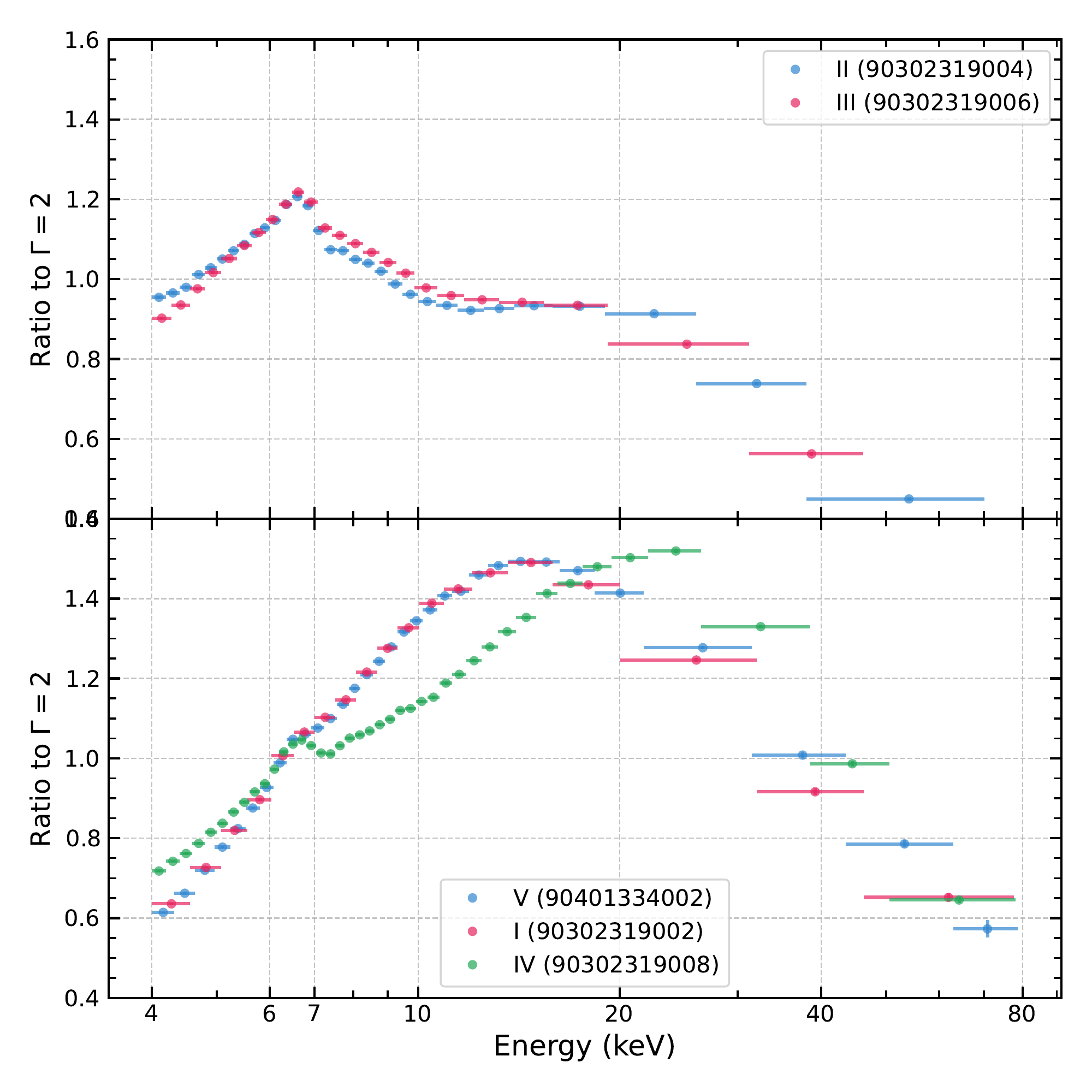}
   \caption{The ratio of the \nustar{}  spectra to the \texttt{powerlaw} model  ($\Gamma=2$) for  observations II and III (upper panel) and  observations I, V and IV (lower panel). Only FPMA data are shown.}
    \label{fig:cutoffpl_ratio_ph_ave}
\end{figure}

A remarkable feature of the spectrum of \obj\ in the super-Eddington state (observations II, III, IV, the 4--79~keV luminosity of the order of $\sim 10^{39}$~\ergps) is the presence of strong reflection features, most notably of the fluorescent line of iron. This is illustrated by  Fig.~\ref{fig:cutoffpl_ratio_ph_ave} where the ratio of the \nustar{} spectra to a simple power-law model with photon index $\Gamma=2$  are shown. For these observations, an approximation of the spectrum with the power-law with an exponential cutoff continuum with a superimposed Gaussian line gives the equivalent width of the line  $0.9\pm0.05$~keV  for the brightest observation. Also quite obvious in the ratio plot are a smeared absorption edge in the $\sim 7-15$~keV range and a  possible Compton reflection hump at the $\sim 15-30$~keV energy domain. All these features are familiar signatures of the reflected emission complicated by photo-ionization effects and, possibly, by relativistic smearing \citep{Ross2005}. 

On the contrary, in the sub-Eddington state (observations I and V) where the source luminosity was $\sim {\rm few}\times 10^{37}$~\ergps{} these features are much less pronounced (Fig.~\ref{fig:cutoffpl_ratio_ph_ave}). Indeed, the equivalent width of the Gaussian line in these observations was $90\pm5$~eV, close to typical values for X-ray binaries in the hard spectral state. During these observations, the source spectrum was quite close to a typical spectrum of an  accreting X-ray pulsar \citep{Jaisawal2018}. 

In observation IV the source was at the declining part of the light curve (Fig.\ref{fig:lightcurve}) and its luminosity was the lowest among the three super-Eddington state observations, $L_{\rm X}~\sim~4\cdot~ 10^{38}$~\ergps. Interestingly, in this observation, the equivalent width of the iron line is large, $0.83\pm0.02$~keV, however, the shape of the continuum  is close to sub-Eddington state spectra found in observations I and V (lower panel in Fig.~\ref{fig:cutoffpl_ratio_ph_ave}). It appears that  this observation caught the source  in  transition from super-Eddington to the sub-Eddington state.

Motivated by the ratio plot in Fig.~\ref{fig:cutoffpl_ratio_ph_ave} we use the following spectral model to describe the emission of the \obj\ in the super-Eddington regime: \texttt{phabs}*(\texttt{relxilllp}+ \texttt{bbodyrad}). The \texttt{relxilllp}\footnote{\url{http://www.sternwarte.uni-erlangen.de/~dauser/research/relxill/}} \citep{Garcia2014, Dauser2016} [v1.4.3] component models reflection of the the primary emission (a power law with exponential cutoff) from an accretion disc in lamp-post geometry, with account for possible photo-ionization of the disc material and relativistic smearing of the reflection spectrum.   Previously, such a  model, in its \texttt{relxill} variant, was successfully applied to the \textit{NICER}+\nustar{} spectra of \obj{} by \citet{Jaisawal2019}. The multiplicative component \texttt{phabs} takes care for the interstellar absorption. The total Galactic absorption in the direction of the source  has the column density of $0.8\times10^{22}$~atoms~cm$^{-2}$  \footnote{\url{https://heasarc.gsfc.nasa.gov/cgi-bin/Tools/w3nh/w3nh.pl}, see \citet{HI4PICollaboration2016}}. For the sake of fit stability, we fix \texttt{phabs} parameter at this value. The \texttt{bbodyrad} component  accounts for thermal radiation reported in previous works.

\begin{figure}
    \centering
    \includegraphics[width=0.5\textwidth]{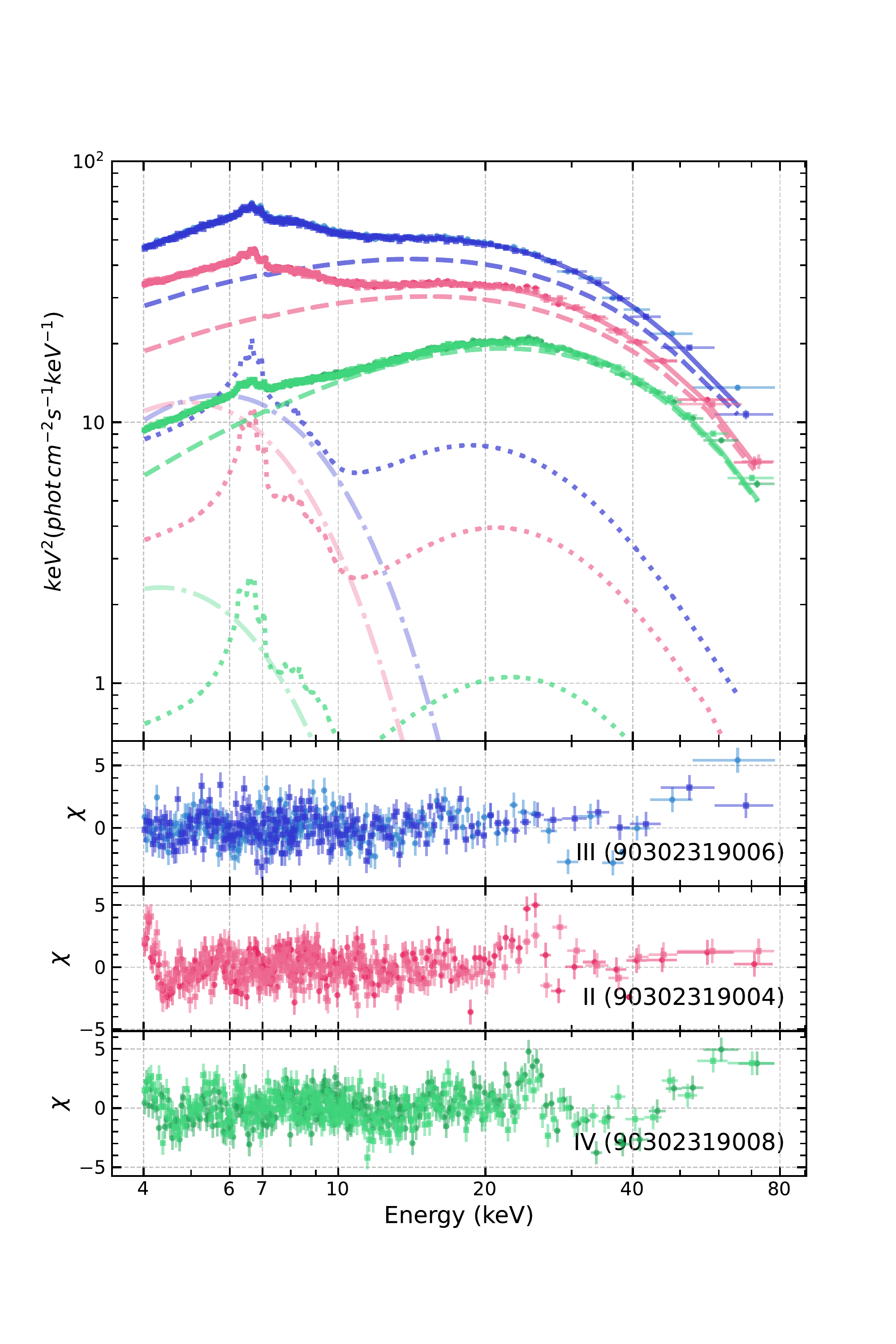}
    \caption{Unfolded spectra of the source and its model for observation II-IV. The top subplots show the unfolded spectra from FPMA (circles) and FPMB (squares) which are coloured according to the observation ID.  A solid line represents the best-fitting approximation, whose components are also  shown: reflection (dotted), power-law continuum (dashed), the thermal radiation (dash-dotted). The lower subplots show the respective residuals from the model. Spectral data were rebinned for plotting purposes.}
    \label{fig:ph-ave}
\end{figure}

For \texttt{relxilllp}, we use the following setup: spin of the object $a$ and its redshift $z$ are both set to $0$; iron abundance $A_{\rm Fe}$ is set to 5.0 \citep{Jaisawal2019}; inner disc radius $R_{\rm in}$,  power law index $\Gamma$, cutoff energy $E_{\rm cut}$ and ionization parameter $log\xi$ are free parameters.  Outer disc radius was set to $1.1\times R_{\rm in}$, this choice will be discussed later. Inclination is a free parameter.
Height of the lamp-post source $h$ is set to $5~R_{\rm g}$ (five gravitational radii).  Reflection factor $f_{\rm refl}$ controls the amplitude of the reflected component relative to the primary continuum and is a free parameter. $f_{\rm refl}$ is an analogue to the equivalent width of the iron line. The cross-calibration constant is added between FPMA and FPMB spectra. Table~\ref{tab:ph-ave} shows the best-fitting values of parameters in  observations II, III, and IV.  Fig. ~\ref{fig:ph-ave} shows the spectral data and models. For observations I and V, when the source was in the sub-Eddington state, the spectra are  adequately  approximated by more simple standard pulsar models \citep{Jaisawal2018} without the need for a significant reflection component. Therefore we focus our attention on the super-Eddington state spectra.

For the super-Eddington state, the reflection model gives fairly consistent values of parameters between three observations.  The spectra in the super-Eddington state require large values of reflection fraction $f_{\rm refl}\sim 0.1-0.2$ which is not surprising, given the large values of the equivalent width of the iron line obtained in  the simple Gaussian fits. Interestingly, the data require both relativistic smearing of the iron line as well as photo-ionization of the reflector's material with a fairly large ionization parameter $\log \xi \sim 3.5$ with a typical uncertainty $0.05$. The inner radius of the disc $R_{\rm in}\sim 50-70~R_{\rm g}$ corresponds to a circular velocity of approximately $10\%$ of the speed of light for a Keplerian disc.  
This explains the significant width of the 6.4~keV line reported earlier \citep{Tao2019, Jaisawal2019} as well as the presence of the lines of He- and H-like iron. These findings are quite important and will be discussed in the context of the truncated thick disc picture in the following section. The best-fitting model consistently  requires  fairly low values of the inclination angle of $\approx10-20\degr$ for all three observations with the errors in $2-4 \degr$ range. However, it is worth noting that in the geometry  proposed in this paper the inclination angle does not have the same straightforward meaning as in the  geometry of a lamp post above an accretion disc envisaged in the {\tt relxilllp} model. This is further discussed in Section 4. 

  As was noted above, in observation IV the source appeared to be  in transition from super- to sub-Eddington state. This may explain the harder photon index and smaller reflection amplitude found in this observation (Table~\ref{tab:ph-ave}). However, the presence of the reflected component in this spectrum is required with high statistical confidence. Likewise, the relativistic broadening of the iron line, expressed in the value of $R_{\rm in}$ parameter, is similar to the other two super-Eddington observations.

Further insights into the  geometry of the super-Eddington accretion flow around the magnetized neutron star are provided by the results of  pulse phase-resolved spectroscopy.

\begin{table*}
    \centering
\begin{tabular}{ll|lll}
\hline
                            &             & II (90302319004)          & III (90302319006)      & IV (90302319008)       \\
Component                        & Parameter         &                           &                        &                        \\ \hline
constant                    & $C_{FPMA/FPMB}$      & $0.999\pm0.001$           & $0.992\pm0.002$        & $0.999\pm0.001$        \\\hline
\multirow{7}{*}{relxilllp} & $E_{\rm cut}$, keV        & $20.8\pm0.5$              & $20.5\pm0.5$           & $17.6^{+0.2}_{-0.3}$   \\
                            & Incl, $\degr$        & $18^{+2}_{-3}$            & $10^{+3}_{-4}$         & $16^{+3}_{-2}$         \\
                            & $R_{\rm in}$, $r_{\rm g}$         & $66\pm6$                  & $53^{+6}_{-5}$         & $51\pm4$               \\

                            & Photon index       & $1.27\pm0.03$             & $1.32\pm0.03$          & $0.77\pm0.02$          \\
                            & log$\xi$       & $3.47^{+0.04}_{-0.05}$    & $3.68^{+0.04}_{-0.03}$ & $3.37^{+0.04}_{-0.07}$ \\
                            & norm        & $0.45\pm0.01$             & $0.65\pm0.02$          & $0.188\pm0.002$        \\
                            & $f_{\rm  refl}$ & $0.184^{+0.011}_{-0.010}$ & $0.24\pm0.01$          & $0.094\pm0.009$        \\\hline
\multirow{2}{*}{bbodyrad}   & $kT$, keV          & $1.223\pm0.008$           & $1.44\pm0.02$          & $1.07\pm0.03$          \\
                            & norm        & $1112^{+52}_{-56}$        & $612^{+91}_{-83}$      & $374^{+26}_{-20}$      \\\hline
      $\chi^2/dof$                  &         & 2545.53/2309              & 2290.51/2170           & 3224.41/2735           \\
$L_{4-79}$, $10^{38}$ erg s$^{-1}$                       &     &         $7.8$                            & $10.1$                           & $3.8$                                               \\ \hline
\end{tabular}
    \caption{Spectral parameters of \nustar{} spectra of \obj{} in its 2017-2018 outburst for super-Eddington state. The spectral model is described in sect.~\ref{results-ph-ave}: \texttt{phabs}*[\texttt{relxilllp} + \texttt{bbodyrad}]. Luminosity is calculated from unabsorbed flux of  \texttt{cflux} component in {\sc xspec}. Black body component normalisation $norm$ is tied to the radius of emitting areas $R$ (in km) through the distance to the source $D$ (in 10 kpc): $norm=R^2/D^2$.  }
    \label{tab:ph-ave}
\end{table*}

\subsection{Phase-resolved spectral analysis}

\label{results-ph-res}

Using the data of three \nustar{} observations of the source in the super-Eddington state, we study  the evolution of spectral parameters as a function of the rotational phase of the neutron star. To this end, the pulse period is divided into 10 phase bins of equal width, and the spectra are extracted for each phase bin as described in section~\ref{nu_data}. We fit the spectra with the same model as for the phase-averaged case, except that we fix $R_{\rm in}$, $\log\xi$, inclination angle,  and parameters of the black body component at the values found for  phase-averaged spectra. For all observations and in phase bins we obtained adequate quality of the fit with the reduced $\chi^2$ in the $\chi^2/{\rm dof}\sim 1.1$ range  for a typical number of degrees of freedom $\sim 1100-1600$. The  phase dependence of  the best-fitting parameters is shown in Fig.~\ref{fig:ph-res-appendix},  and the representative models in Fig.~\ref{fig:ph-res-spe-appendix}. The parameters value are presented in Table \ref{tab:ph-res}.   Making $\log\xi$ and parameters of the black body component free  does not change significantly other parameters of the model and does not affect the interpretation of the results. For simplicity, we do not consider these more sophisticated fits further and focus on our  main finding of the presence of a strong reflected component and its behaviour with the pulse phase.

\begin{figure}
    \centering
    \includegraphics[width=0.5\textwidth]{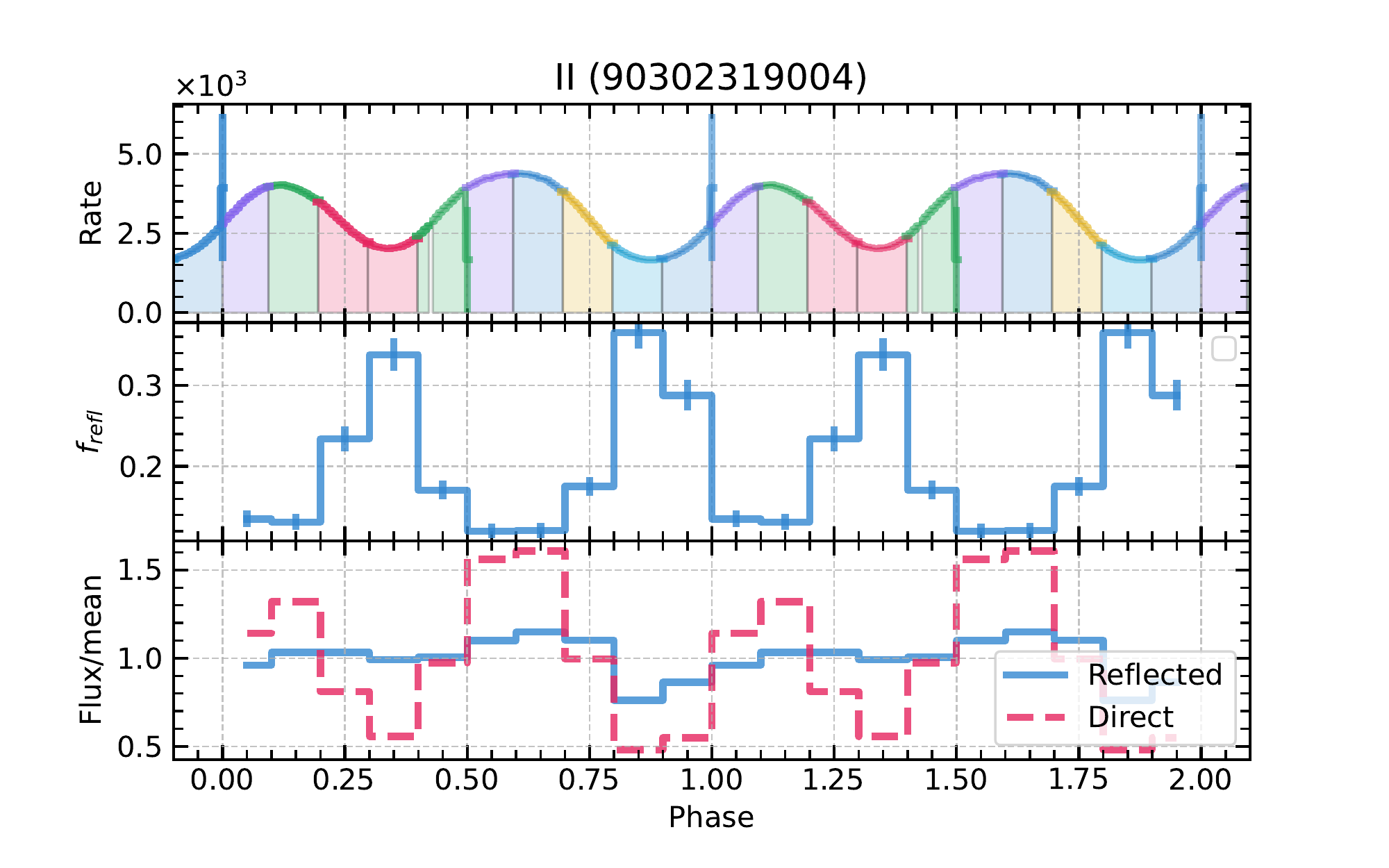}
    \includegraphics[width=0.5\textwidth]{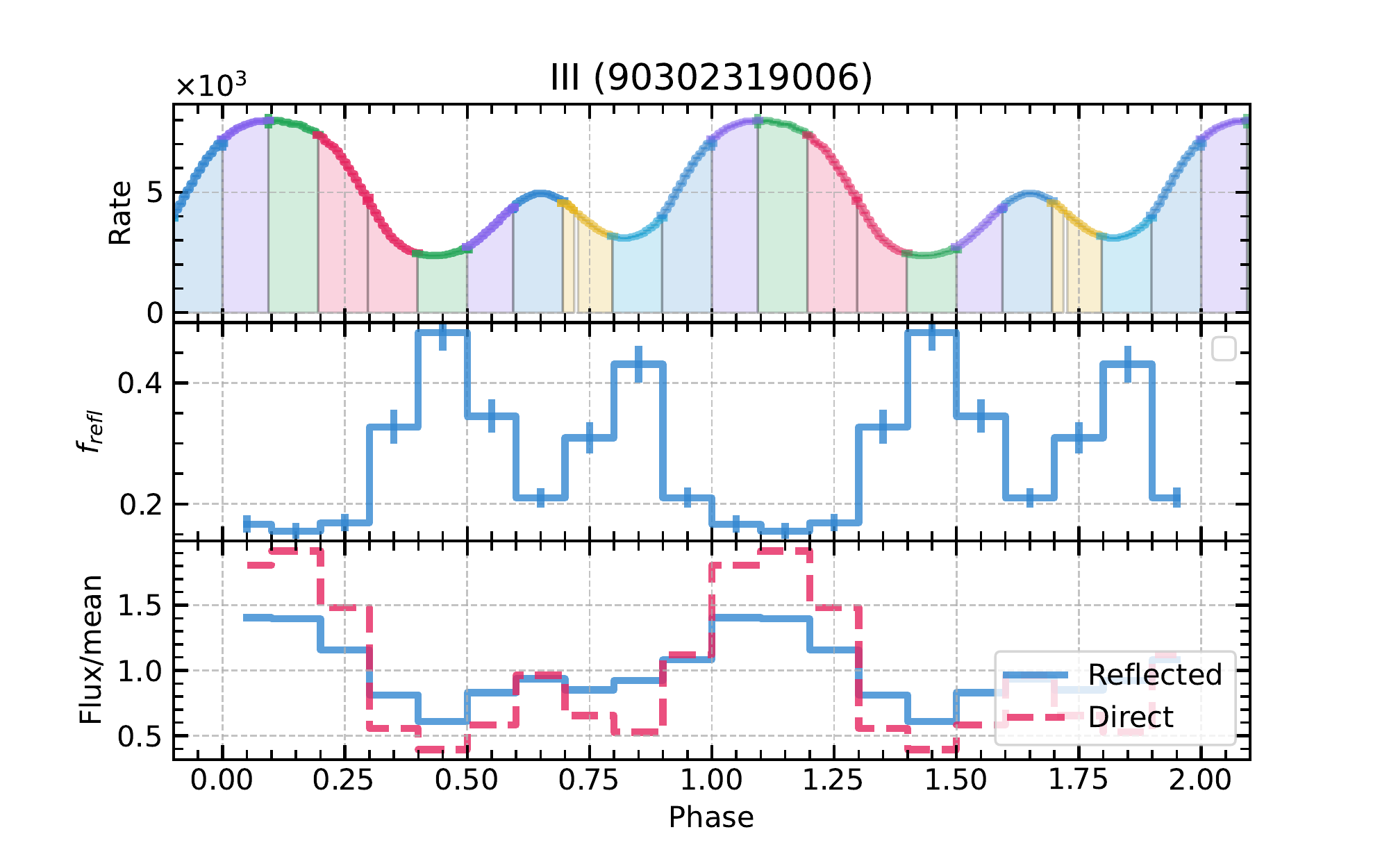}
    \includegraphics[width=0.5\textwidth]{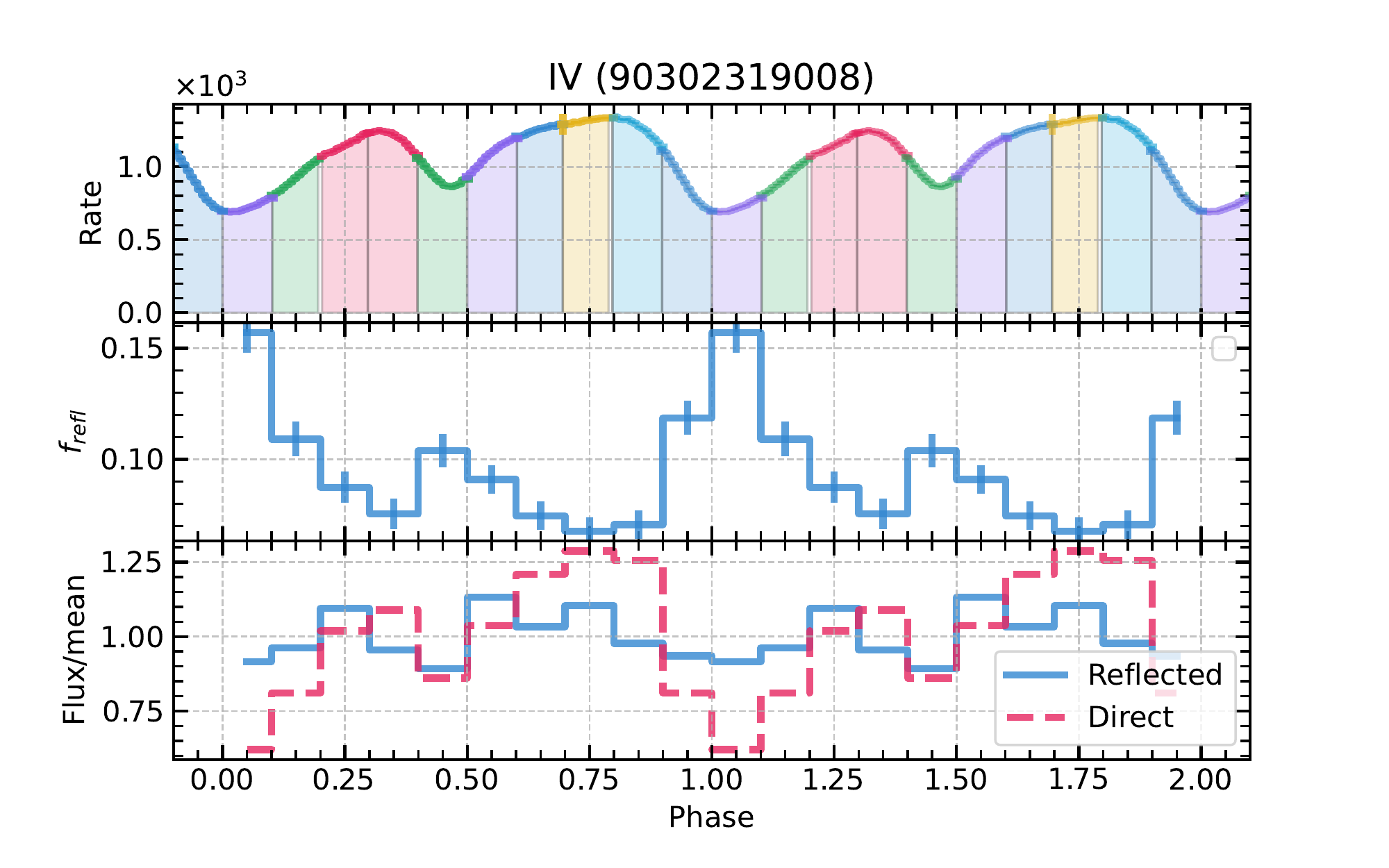}
    \caption{The evolution of spectral parameters as the function of pulsar rotation phase for observations II, III, and IV. In each panel from top to bottom subplot: Pulse profile of the count rate in the 4--79~keV energy range; the reflection fraction parameter $f_{\rm refl}$; direct (red, dashed) and reflected(blue, solid) flux (4--50~keV) of the  best-fitting model divided by the respective mean flux. All spectral parameters may be found in Fig.~\ref{fig:ph-res-appendix}}

    \label{fig:ph-res}
\end{figure}

Using the best-fitting model we compute the flux of reflected and direct emission for each spectrum. We do not compute the uncertainties in those values since it would require computationally-heavy calculations  of the parameter chains which is unnecessary because the errors are sufficiently small for the purpose of this work. This is demonstrated by the high accuracy of $f_{\rm refl}$ estimates as well as by the low level of noise in the obtained pulse profiles for the direct and reflected flux. Phase dependence of the direct and reflected flux as well as the total flux and reflection fraction $f_{\rm refl}$ are shown in Fig.~\ref{fig:ph-res}. An immediate conclusion from these plots is that the reflected component is much less variable with the pulse phase than the direct emission. As discussed in the next section, this can be understood in a picture of a rotating neutron star located in the middle of a geometrically thick super-Eddington accretion disc truncated by the magnetic field of the strongly magnetized neutron star.  The pulsation amplitude\footnote{i.e. $\frac{F_{\rm max}}{F_{\rm min}}$} of the direct emission is $\approx 3.5, 5.0, 2.0$ in observations II, III and IV respectively, whereas for the reflected emission the numbers are, respectively, $\approx 1.5, 2.3, 1.3$. The pulsation amplitude for the reflected emission is $\sim 2-3$ times smaller than for the direct emission, which leads to the large pulsations in the reflected fraction $f_{\rm refl}$.

Other parameters of the model show a complex pattern of variations with the pulse phase (Fig.~\ref{fig:ph-res-appendix}). The spectral hardness and the cut-off energy seem to be slightly correlated  with the total pulsar flux, as commonly observed in such objects \citep[e.g.][]{Tsygankov2017}.

The results of the phase resolved spectral analysis are summarised  in Table \ref{tab:ph-res}. In interpreting  values of the photon index one should bear in mind that in the relxill model, the direct emission is modelled as a power law with exponential cut-off. Typical values of the cut-off energy $E_{\rm cut}$ for this sources are in the 15-30 keV range. Therefore the actual slope of the spectrum is not well represented by solely the value of the slope $\Gamma$, but, rather, by combination of Gamma and $E_{\rm cut}$. Variations of the slope with pulse phase  are accompanied by corresponding variations of the $E_{\rm cut}$ so that drastic variations of $\Gamma$ and $E_{\rm cut}$ from one pulse phase bin to the other do not mean equally drastic variations of the shape of the continuum.

\begin{figure}
    \centering
    \includegraphics[width=0.5\textwidth]{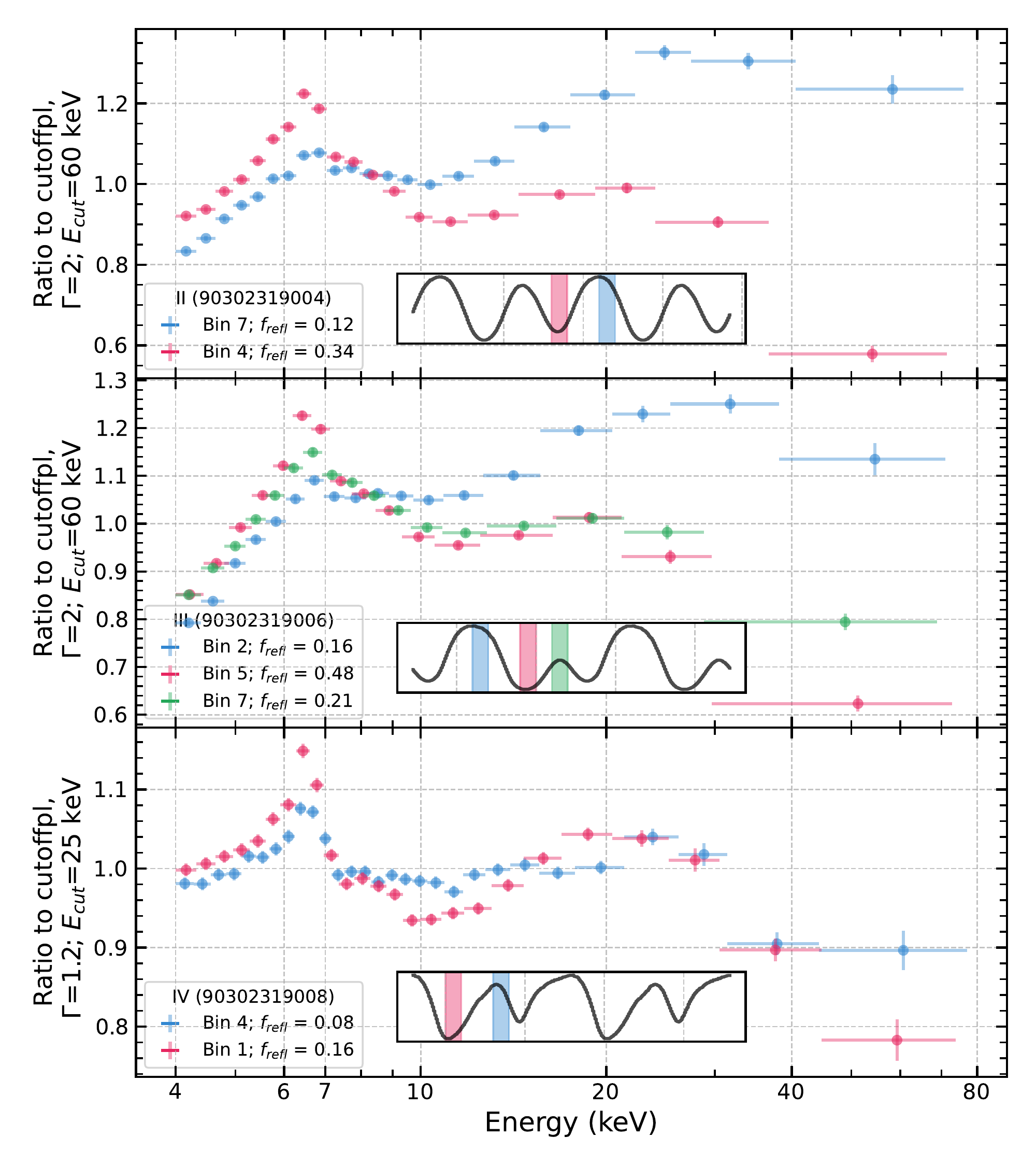}
    \caption{The ratio of the spectra collected in specified phase intervals to \texttt{cutoffpl} model with fixed parameters. The colour of the spectra corresponds to the colour of  a phase bin shown in the pulse profile inset. Data from both detectors are plotted in the figure (square and circle markers). The top panel shows the results for observation II, the middle panel for observation III, and the bottom for observation IV. In the legend of each panel, the value of $f_{\rm refl}$ from the best-fitting model of a bin is shown. Only FPMA data are shown.}
    \label{fig:ph-res-spe-rat}
\end{figure}

The presence of variations in the strength of the reflected component with the pulsar phase is further illustrated in Fig.~\ref{fig:ph-res-spe-rat} where the ratio of the spectra in different phase bins to a {\tt cutoffpl} model with the same parameters is shown for several phase bins.
One can see that the amplitude of the reflection features is indeed larger in the spectra for which the fit gave larger values of $f_{\rm refl}$.

We find  no  evidence of the presence of the cyclotron scattering absorption feature (CSRF) in phase-averaged and phase-resolved spectra in the 4--79 keV energy range. 

We tried both lamp-post (relxilllp) and corona (relxill) models with no noticeable effect on the parameter values. The lamp-post geometry was motivated by the compactness of the accretion column compared to the inner disk radius that we envisaged.
We also varied the lamp-post height, tried fixing it to 3, 5, 10 and 40 $R_g$ and found that it did not have any effect on the best-fit values of the parameters except for the normalisation of the the relxillp model. Notably, $f_{\rm refl}$ and cutoff energy/photon index did not change. It  should be mentioned that both lamp-post and corona models are a significant simplifications of the true geometry of the system. However the agreement between the results we obtained under these geometries suggest that the exact geometry is not the critical parameter.

\section{Discussion}
\label{disussion}

The presence of pulsations, i.e. lack of the propeller regime \citep{Illarionov1975} down to the luminosity level of  $\sim3\times10^{34}$~\ergps{} found by \citet{Doroshenko2020} implies that the dipole component of the magnetic field is not very strong. Based on this and a few other arguments  \citet{Doroshenko2020} estimated the dipole field value of the pulsar $B\sim(3-9)\times 10^{12}$~G and coupling constant $\xi\sim0.1-0.2$ \footnote{the $\xi$ is defined as a parameter relating  the magnetospheric radius $R_{\rm m}$ and the Alfven radius $R_{\rm a}$ via $R_{\rm m}=\xi R_{\rm a}$, see, e.g., \citet{Mushtukov2017}}. 
For such field strengths and super-Eddington luminosity of $L_{\rm X}\approx10^{39}$~\ergps{} at the peak of the light curve  one may expect that the pulsar magnetosphere is fairly compact, with  $R_m\sim 150-500$~km, i.e.  $R_m\sim 35-120 R_{\rm g}$ for a 1.4~$M_\odot$ neutron star \citep{Mushtukov2017, Tsygankov2018}. 

Given a rather small  truncation radius, the disc will extend sufficiently close to the compact object, so that a radiation pressure dominated geometrically thick disc can form near the pulsar magnetosphere in the super-Eddington regime. Using formulae from \citet{Shakura1973,  Mushtukov2017}  one may estimate the $H/R$ at the inner edge of the disc  ${\left(\frac{H}{R}\right)_m  = 0.1 \frac{L_{\rm X}}{10^{39}} \frac{R_{\rm NS}}{10^6 \text{cm}} \left(\frac{M_{\rm NS}}{M_{\odot}}\right)^{-1} \left(\frac{R_m}{10^8 \text{cm}}\right)^{-1}}$, with the parameters normalized to values appropriate for  \obj. Substituting the values, one obtains $H/R\sim0.15-0.5$ on the inner edge of the disc.  
Indeed, \citet{Doroshenko2020}, based on the observed transitions in spectral and timing properties of the source, argued that between MJD~58045 and 58098 the inner part of the accretion disc was in the radiation pressure dominated regime. All three observations analysed in the present paper are within this time span.

\begin{figure}
    \centering
    \includegraphics[width=0.5\textwidth]{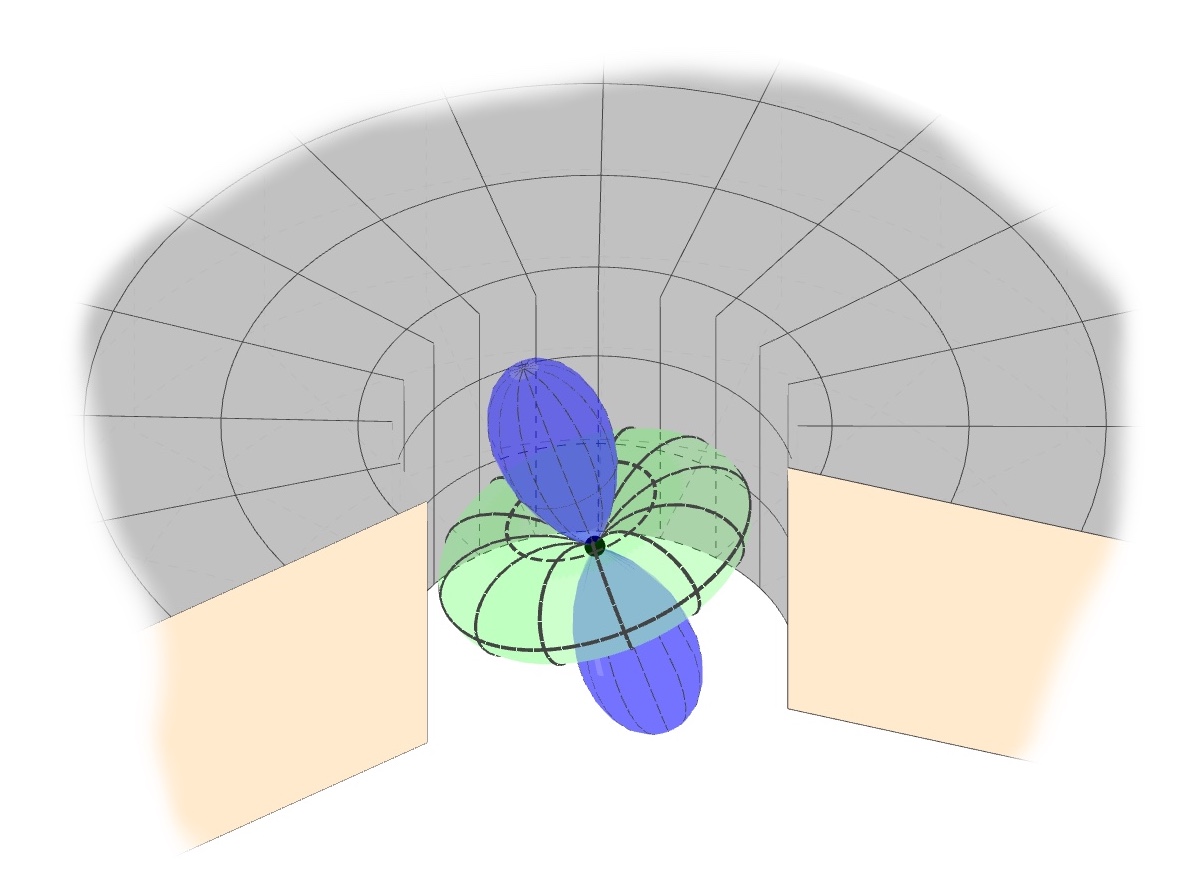}
    \caption{The illustration of considered geometry: a \textbf{neutron star} (black dot) surrounded by a \textcolor{gray}{\textbf{thick disc}} (gray  plane) illuminated by a  \textcolor{blue}{\textbf{pencil  beam}} (blue cone), and  the \textcolor{green}{\textbf{fan beam}} (green torus). Depending on the orientation of the pulsar and its emission diagram, the visible side of the inner disc may be illuminated differently depending on the rotational phase.}
    \label{fig:pulsar}
\end{figure}

Truncation of the geometrically thick accretion disc by the magnetic field of the neutron star will form a "well" with the pulsar located in its centre (see the sketch in Fig.~\ref{fig:pulsar}). The inner edge of the "well" will be subject to continuous illumination by  X-ray emission from the pulsar \citep{Mushtukov2021}, similar to illumination of the accretion disc in X-ray binaries and active galactic nuclei at more "normal" sub-Eddington  accretion rates. However, unlike X-ray binaries at sub-Eddington accretion rate, the solid angle subtended by the inner edge of the thick accretion disc with $H/R\approx 1$ would be much larger. The solid angle of the inner edge of a disc with aspect ratio $H/R=1$  is $2\pi\int_{\frac{\pi}{2} - \operatorname{arctan}(H/R)}^{\frac{\pi}{2} + \operatorname{arctan}(H/R)}\operatorname{sin}\theta \operatorname{d}\theta =0.7\times4\pi$~sr. Thus, $\Omega/4\pi\sim 0.7$ and the ballpark figure for equivalent width of the fluorescent line of iron  is $\sim \frac{\Omega}{4\pi}\times1$~keV $\sim0.7$~keV assuming reflection from the optically thick material. This is compatible with the observed equivalent width of the 6.4~keV line of  the order of 1~keV  \citep[see sect. \ref{results-ph-ave}, also][]{Tao2019}.
Similarly, the value of $f_{\rm refl}\sim 0.25$ obtained in the spectral fits corresponds to $\Omega/4\pi\sim 0.2$\footnote{according to the \texttt{relxilllp} model, $f_{\rm refl}\sim \frac{\Omega}{4\pi - \Omega}$}, which is comparable with the value estimated above. We note however, that this crude estimate assumes an isotropic emission pattern, which is not strictly applicable to a magnetized neutron star.

Furthermore, we note that our best-fitting model predicts the inner radius of the accretion disc $\sim50-70~R_{\rm g}\sim (2-3)\times10^7$ cm, with the typical uncertainty of 5 $R_{\rm g}$. This value is controlled by the line profiles  in the iron line complex and was determined in the \texttt{relxilllp} model with the account for the Doppler and  ionization effects (both of which contribute to the line broadening). This value of the inner disc radius   estimated from the  broadening of the line is consistent with the radius of the pulsar magnetosphere  $\sim 3\times10^7$~cm estimated  assuming the (dipole) magnetic field strength of $B=3\times10^{12}$~G and the coupling parameter of $\xi=0.2$ and   \citep[eq. 1 in][]{Mushtukov2017}. This consistency is quite remarkable because the two estimates are based on entirely different physical effects (Doppler-effect and disc-magnetosphere interaction) and are independent. 

Thus, we  hypothesise that the strong reflected emission originates at the inner edge of the thick accretion disc truncated by the magnetic field of the neutron star. The inner edge of the disc  is  irradiated by hard X-ray  emission of an accretion column, as illustrated in  Fig.~\ref{fig:pulsar}.  The pulsar emission diagram must be sufficiently broad so that the total reflected  flux, as seen by the distant observer, does not vary much as the neutron star rotates. As the neutron star rotates, the pulsar beam  sweeps through the line of sight giving rise to the direct emission pulsating with a much larger amplitude. These two components, weakly variable reflected emission and more strongly variable direct emission from the pulsar explain the overall behaviour of the pulse with strongly varying reflection fraction. In this picture, the anti-correlation of the reflection fraction and equivalent width of the fluorescent line of iron with the total flux is naturally explained.
On the next level of detalization we may note that, as the pulsar emission is not isotropic, the  irradiation pattern varies somewhat as the neutron star rotates, leading to some moderate variations of the reflected flux, which have however  smaller fractional amplitude than the direct emission from the pulsar itself.

At the considered luminosity levels, the emission diagram of the pulsar is likely of the fan-like geometry,  i.e. sufficiently broad, the pencil-beam being subdominant, if any.  In this regime, the radiation escapes a super-critical accretion column mostly via its sides, leading to a fan-like emission pattern \citep{Basko1976,  Lyubarskii1982, Mushtukov2015, Mushtukov2022}. Fan-beam naturally produces weakly variable irradiation of the inner disc edge, due to rather high degree of symmetry  of the fan beam as seen from the accretion wall, and, importantly, two-peaked pulse profiles are observed in this state. This is opposite to pencil-beam geometry (expected at smaller accretion rates) when  photons escape primarily along the magnetic axis of the pulsar \citep{Gnedin1973} and we expect a one-peaked pulse profile and little irradiation \citep{Koliopanos2016}. Variability of the direct emission  is caused by the rotation of the fan- and  pencil-beam emission components,  the latter being sub-dominant at considered luminosities. The suggested picture explains observed behaviour with variable direct emission flux and variable reflection fraction, but significantly less variable  reflected flux.

One can estimate the beaming factor of the pulsar comparing reflected and direct flux as follows. Using observation II as an example, the reflected flux in the average spectrum amounts to $\sim 2 $ (in units of $10^{-8}$~\ergpspcm). X-ray albedo varies depending on the ionization state and incidence angles, but generally, it is greater than $\sim 0.3$ \citep{Basko1974,Zycki1994}. Taking into account X-ray albedo we estimate the effective angle-averaged direct emission required to produce observed reflected emission as $\sim 2/0.3=6$ (in the same units). Observed  direct flux is $12$, which is about $\sim$two times higher than the former figure. The ratio of the two flux values gives an estimate of the beaming factor of the order of $\sim 2$, i.e. beaming of the pulsar emission is rather weak, and the measured luminosity is fairly close to the real one.  Beaming of radiation in  X-ray pulsars is discussed, for example, in
 \citet{King2019, Mushtukov2021}.

Outside the outburst peak, at the sub-Eddington accretion rates, the accretion disc returns to the normal geometrically thin state characterised by a rather low reflection fraction, typical for accreting X-ray pulsars, i.e. iron equivalent width $\sim 100$~eV \citep[e.g.][]{Bykov2021} .

With the advent of X-ray polarimetry instruments (e.g. The Imaging X-ray Polarimetry Explorer, IXPE, \citet{Weisskopf2016}) it may be anticipated that  studies of the variability of polarization and its energy dependence will bring new interesting results.  Reflected (reprocesses) emission is expected to have a smaller degree of polarization than radiation coming from the accretion columns \citep{Basko1975}. The degree of polarization can be used to disentangle the direct pulsar emission from the reflected and/or thermal components, which brings the model-independent way to study the reprocessed emission and hence the matter distribution in such systems.

\section{Conclusion}
\label{conclusion}

We analysed \nustar{} data of the  ultra-luminous X-ray pulsar \obj\ in the super-Eddington state. In this state, the spectrum of the source shows strong reflection signatures (fluorescent lines of iron with the equivalent width of $\sim 1$~keV, smeared K-edge and Compton hump) suggesting that the spectrum is dominated by the reflected emission. Pulse phase-resolved spectral analysis using a reflection model demonstrated significant modulation of the reflection fraction $f_{\rm refl}$. However, a more careful look at the data revealed that this modulation is caused by  variations of the direct emission,  the reflected flux being significantly less variable. 

We hypothesized that due to the large accretion rate and relatively moderate dipole magnetic field strength in this pulsar, the configuration is realised when the geometrically thick radiation pressure dominated accretion disc is truncated close to the neutron star, at the distance of $\sim 50-75~R_{\rm g}$. The inner edge of the truncated accretion disc is filling a large solid angle as seen from the pulsar and is continuously illuminated by the X-ray radiation from the accretion column.  The beaming factor of the emission is not large so that the radiation  has a sufficiently broad fan-like emission diagram, expected at these large luminosities. This gives rise to the strong and weakly variable reflected emission.  As the  emission from the accretion column sweeps through the line of sight, we see  direct emission of the accretion column, which is much more variable. 

From the spectral analysis, based on the width of the iron lines and with the account for the photo-ionization of the material at the inner edge of the accretion disc we estimate its radius to be $\sim 50-70~R_{\rm g}$. This value is remarkably consistent with the magnetospheric radius predicted given the constraints on the neutron star magnetic field strength and coupling parameter.

The uniqueness of this pulsar is explained by the combination of the moderately super-Eddington accretion rate with the  moderate magnetic field strength.

\section*{Acknowledgements}
We thank the anonymous referee for useful and constructive comments and suggestions which helped to improve the presentation of our results.
Sergei Bykov acknowledges support from and participation in the International Max-Planck Research School (IMPRS) on Astrophysics at the Ludwig-Maximilians University of Munich (LMU).  This work was partially supported by the grant 14.W03.31.0021 of the Ministry of Science and Higher Education of the Russian Federation.

Facility: \nustar\, \citep{Harrison2013}.

Software:  NumPy \citep{Harris2020}, Matplotlib \citep{Hunter2007},  SciPy \citep{2020SciPy-NMeth}, AstroPy \citep{astropy:2018}, Pandas\citep{reback2020pandas}, HEASoft \footnote{\url{https://heasarc.gsfc.nasa.gov/docs/software/heasoft/}}.
 Code used to produce the results of the paper would be available shortly after the publication\footnote{\url{https://github.com/SergeiDBykov/nustar_sj0243}}. 
\section*{Data Availability Statement}
We are grateful for the \nustar{} data to the High Energy Astrophysics Science Archive Research Center (HEASARC) Online Service \footnote{\url{https://heasarc.gsfc.nasa.gov/cgi-bin/W3Browse/w3browse.pl}} provided by the NASA/Goddard Space Flight Center. To download data of \nustar{} used in this work, one should use observations ID from sect.~\ref{nu_data}. \textit{Swift} data for light curves are available from the links in sect.~\ref{results-timing}.  



\bibliographystyle{mnras}
\bibliography{sj0243} 



\appendix
\section{Phase-resolved parameters}

\begin{figure*}%
    \centering
    \subfloat[\centering Observation II]{{\includegraphics[width=.45\textwidth]{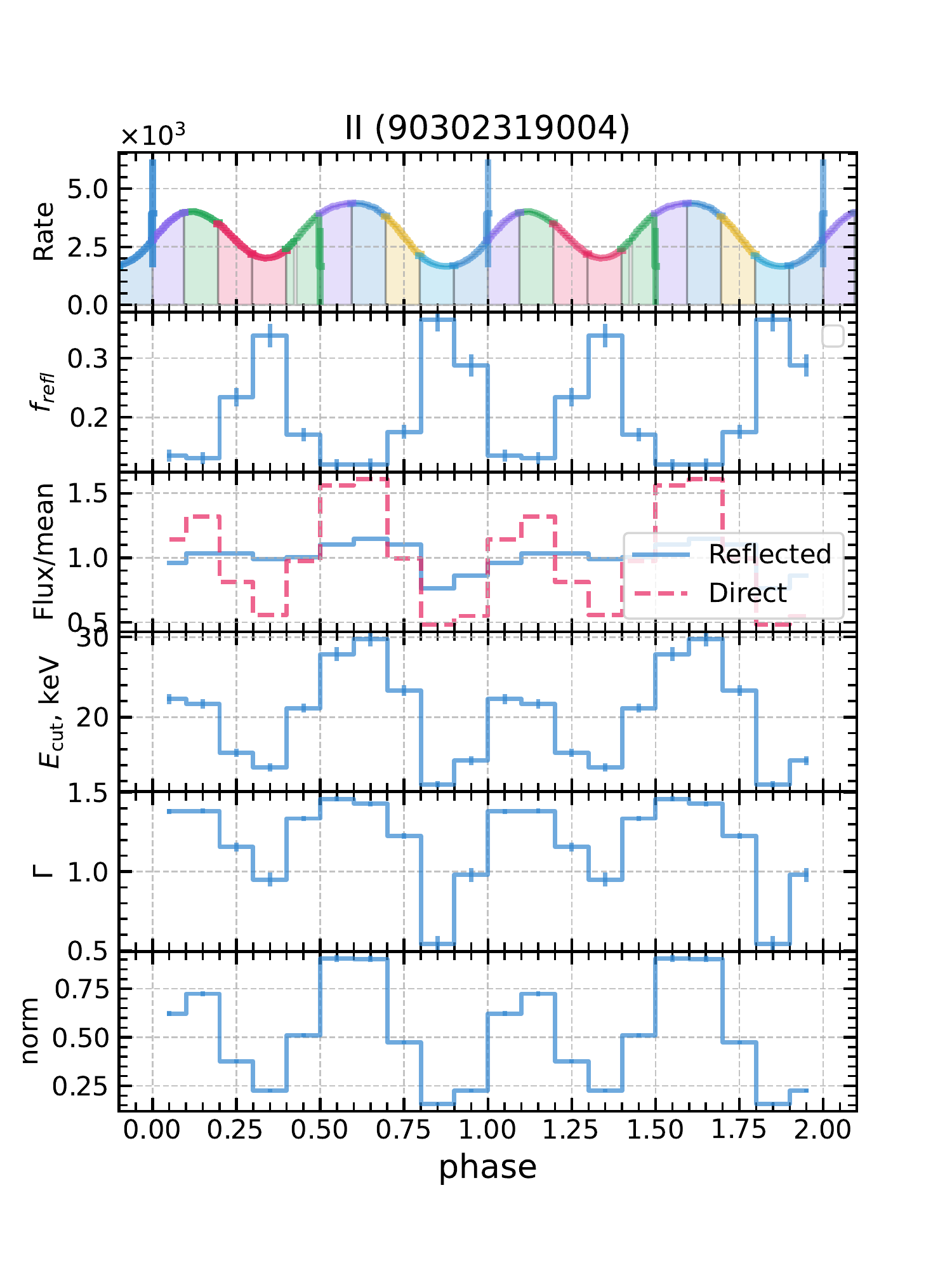} }}%
    \subfloat[\centering Observation III]{{\includegraphics[width=.45\textwidth]{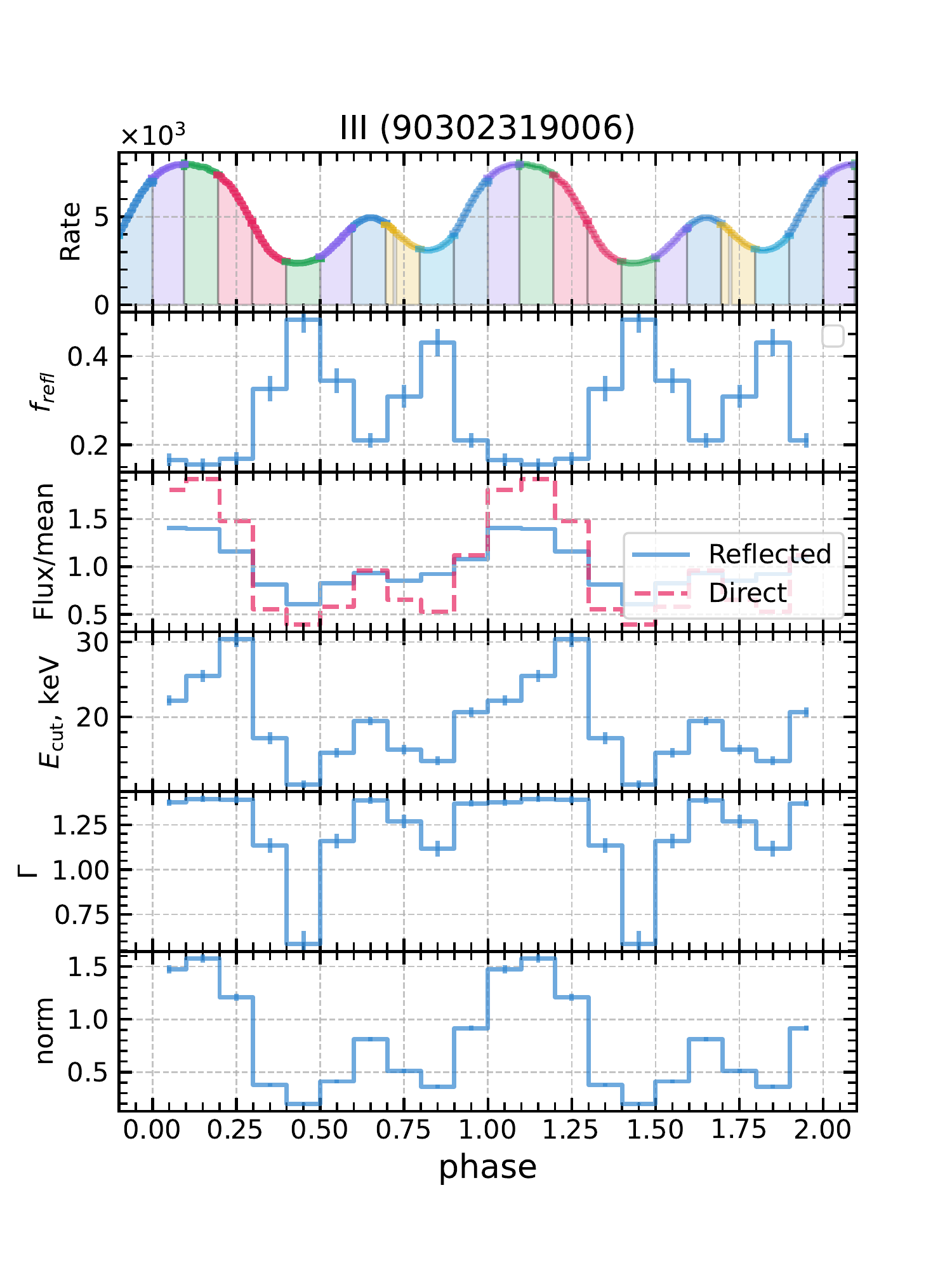} }}%
     \qquad
     \subfloat[\centering Observation IV]{{\includegraphics[width=.45\textwidth]{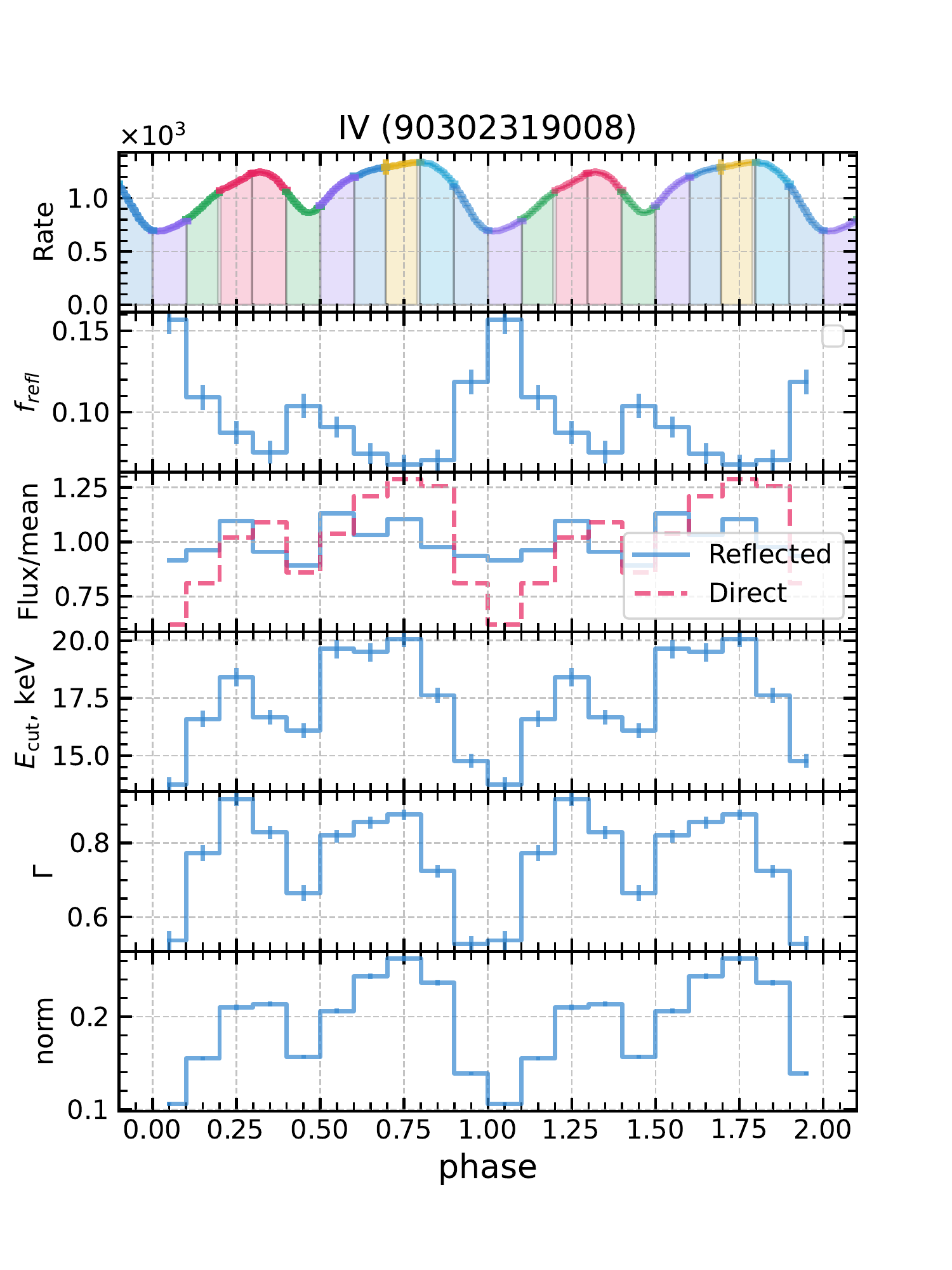}}}%
    \caption{ The evolution of spectral parameters as the function of pulsar rotation phase for observations II, III and IV. In each panel from top to bottom subplot: Pulse profile of the count rate in the 4--79~keV energy range; the reflection fraction parameter $f_{\rm refl}$;  direct (red, dashed) and reflected(blue, solid) flux (4--50~keV) of best-fitting model divided by the respective mean flux; e-folding energy of the continuum $E_{\rm cut}$;  Photon index $\Gamma$;  normalization of \texttt{relxilllp} model.  Two phase intervals are plotted.}   
    \label{fig:ph-res-appendix}%
\end{figure*}

\begin{figure*}%
    \centering
    \subfloat[\centering Observation II]{{\includegraphics[width=.4\textwidth]{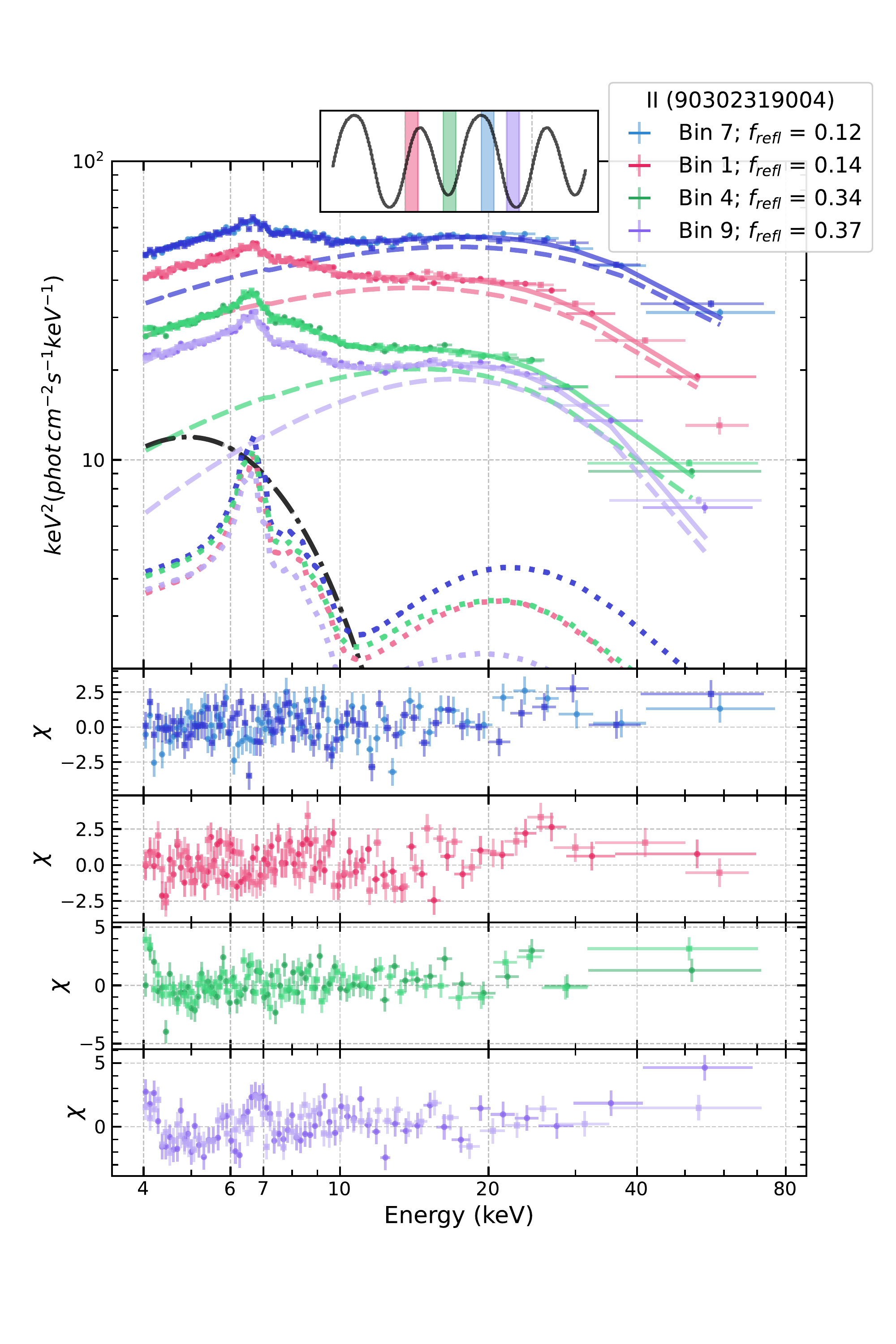} }}%
    \subfloat[\centering Observation III]{{\includegraphics[width=.4\textwidth]{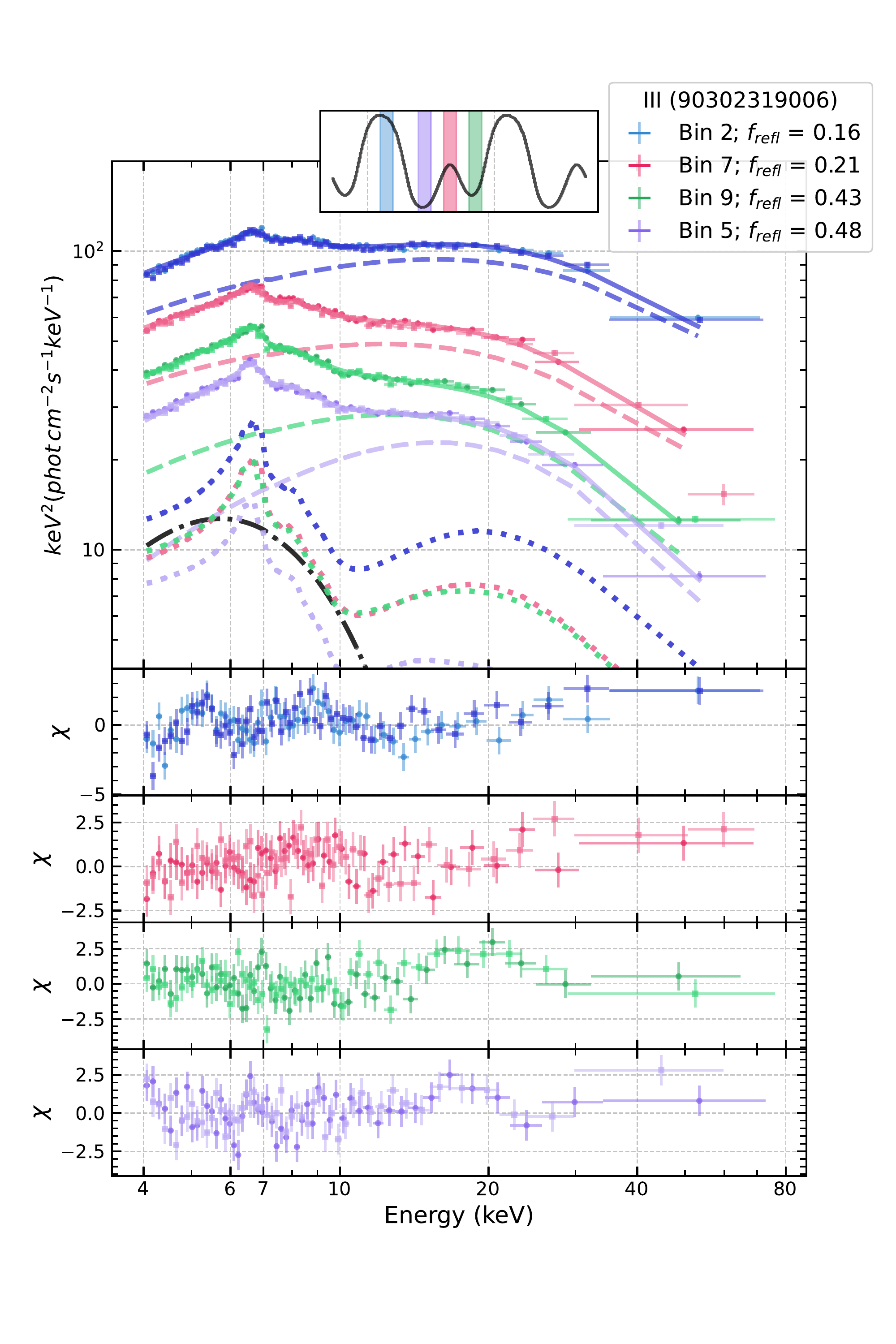} }}%
     \qquad
     \subfloat[\centering Observation IV]{{\includegraphics[width=.4\textwidth]{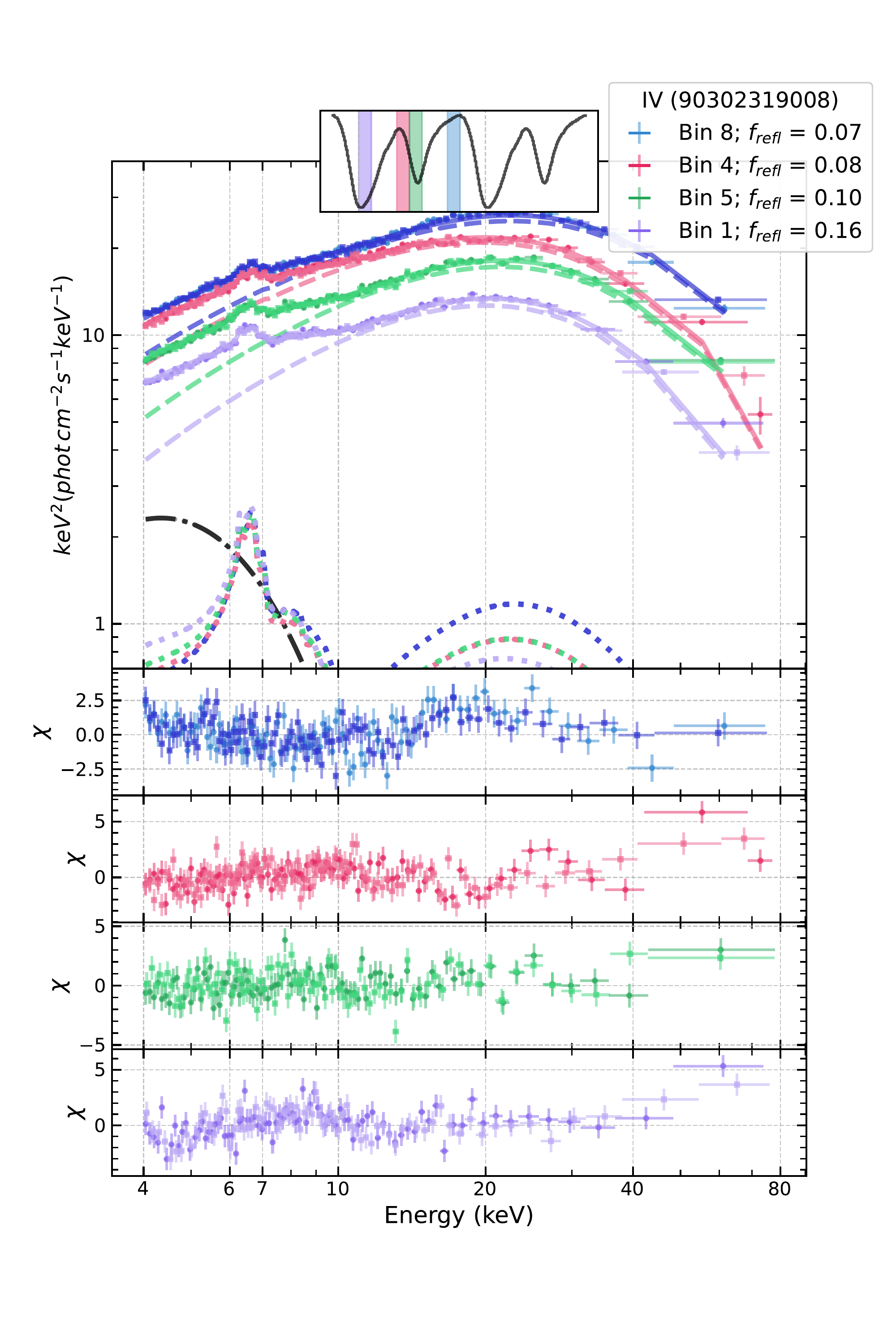}}}%
    \caption{As in Fig.~\ref{fig:ph-ave} but for spectra of different rotational phases in observation  II, III and IV (color of the a spectrum corresponds to the color of a phase-bin in the pulse profile inset). In the legend the reflected fraction $f_{\rm frac}$ of the \texttt{relxilllp} model is shown for a given bin.} 
    \label{fig:ph-res-spe-appendix}%
\end{figure*}

\begin{table*}
    \centering
    \begin{tabular}{ccccccc}
\hline
phase &   $f_{\rm refl}$ & \begin{tabular}{@{}c@{}}$F_{\rm refl}$ \\ $(10^{-8}$ erg cm$^{-1}$s$^{-1}$)\end{tabular}     & \begin{tabular}{@{}c@{}}$F_{\rm dir}$ \\ $(10^{-8}$ erg cm$^{-1}$s$^{-1}$)\end{tabular}  &         $E_{\rm cut}$, keV &       $\Gamma$ &                       norm \\
\hline
II (90302319004) \\

0.05  &    $0.14\pm0.01$ &                                 1.46 &                                13.0 &  $22.3^{+0.7}_{-0.6}$ &  $1.38\pm0.02$ &              $0.62\pm0.01$ \\
0.15  &  $0.131\pm0.010$ &                                 1.57 &                                15.0 &          $21.6\pm0.6$ &  $1.38\pm0.02$ &              $0.72\pm0.01$ \\
0.25  &    $0.23\pm0.02$ &                                 1.57 &                                 9.2 &          $15.5\pm0.5$ &  $1.16\pm0.03$ &              $0.38\pm0.01$ \\
0.35  &    $0.34\pm0.02$ &                                 1.50 &                                 6.3 &          $13.7\pm0.5$ &  $0.95\pm0.04$ &            $0.225\pm0.007$ \\
0.45  &    $0.17\pm0.01$ &                                 1.52 &                                11.1 &          $21.1\pm0.6$ &  $1.34\pm0.02$ &              $0.51\pm0.01$ \\
0.55  &  $0.120\pm0.009$ &                                 1.67 &                                17.7 &          $27.8\pm0.9$ &  $1.46\pm0.02$ &              $0.90\pm0.02$ \\
0.65  &  $0.121\pm0.009$ &                                 1.74 &                                18.3 &          $29.7\pm1.0$ &  $1.43\pm0.02$ &              $0.90\pm0.02$ \\
0.75  &    $0.18\pm0.01$ &                                 1.67 &                                11.3 &  $23.3^{+0.7}_{-0.6}$ &  $1.22\pm0.02$ &            $0.474\pm0.010$ \\
0.85  &    $0.37\pm0.02$ &                                 1.16 &                                 5.5 &          $11.6\pm0.4$ &  $0.54\pm0.05$ &  $0.157^{+0.005}_{-0.004}$ \\
0.95  &    $0.29\pm0.02$ &                                 1.31 &                                 6.2 &  $14.6^{+0.6}_{-0.5}$ &  $0.98\pm0.04$ &            $0.225\pm0.007$ \\
\hline
III (90302319006) \\

0.05  &  $0.17\pm0.01$ &                                 4.92 &                                30.9 &          $22.2\pm0.7$ &           $1.37\pm0.02$ &    $1.47\pm0.04$ \\
0.15  &  $0.16\pm0.01$ &                                 4.88 &                                32.8 &          $25.5\pm0.8$ &           $1.39\pm0.02$ &    $1.57\pm0.04$ \\
0.25  &  $0.17\pm0.01$ &                                 4.06 &                                25.3 &              $30\pm1$ &           $1.39\pm0.02$ &    $1.21\pm0.03$ \\
0.35  &  $0.33\pm0.03$ &                                 2.85 &                                 9.5 &          $17.2\pm0.8$ &           $1.14\pm0.04$ &    $0.38\pm0.02$ \\
0.45  &  $0.48\pm0.03$ &                                 2.13 &                                 6.7 &          $11.0\pm0.5$ &           $0.58\pm0.08$ &  $0.198\pm0.008$ \\
0.55  &  $0.34\pm0.03$ &                                 2.90 &                                10.0 &  $15.2^{+0.7}_{-0.6}$ &           $1.16\pm0.04$ &    $0.41\pm0.02$ \\
0.65  &  $0.21\pm0.02$ &                                 3.28 &                                16.5 &  $19.5^{+0.6}_{-0.5}$ &           $1.39\pm0.02$ &    $0.81\pm0.02$ \\
0.75  &  $0.31\pm0.03$ &                                 2.98 &                                11.2 &  $15.7^{+0.7}_{-0.6}$ &           $1.27\pm0.04$ &    $0.51\pm0.02$ \\
0.85  &  $0.43\pm0.03$ &                                 3.23 &                                 9.0 &          $14.2\pm0.6$ &  $1.12^{+0.04}_{-0.05}$ &    $0.36\pm0.02$ \\
0.95  &  $0.21\pm0.02$ &                                 3.79 &                                19.2 &          $20.7\pm0.6$ &           $1.37\pm0.02$ &    $0.92\pm0.03$ \\
\hline
IV (90302319008)\\

0.05  &            $0.157\pm0.009$ &                                 0.33 &                                 3.7 &          $13.8\pm0.3$ &           $0.54\pm0.03$ &            $0.106\pm0.002$ \\
0.15  &            $0.109\pm0.008$ &                                 0.35 &                                 4.8 &  $16.6^{+0.4}_{-0.3}$ &           $0.77\pm0.02$ &            $0.155\pm0.002$ \\
0.25  &            $0.087\pm0.007$ &                                 0.40 &                                 6.1 &          $18.4\pm0.4$ &           $0.92\pm0.02$ &            $0.210\pm0.003$ \\
0.35  &            $0.075\pm0.007$ &                                 0.34 &                                 6.5 &          $16.7\pm0.3$ &           $0.83\pm0.02$ &            $0.213\pm0.003$ \\
0.45  &  $0.104^{+0.008}_{-0.007}$ &                                 0.32 &                                 5.1 &          $16.1\pm0.3$ &           $0.67\pm0.02$ &            $0.157\pm0.002$ \\
0.55  &  $0.091^{+0.007}_{-0.006}$ &                                 0.41 &                                 6.2 &          $19.7\pm0.4$ &  $0.82^{+0.01}_{-0.02}$ &            $0.206\pm0.003$ \\
0.65  &            $0.074\pm0.006$ &                                 0.37 &                                 7.2 &          $19.5\pm0.4$ &  $0.86^{+0.01}_{-0.02}$ &  $0.244^{+0.003}_{-0.004}$ \\
0.75  &            $0.068\pm0.006$ &                                 0.40 &                                 7.7 &          $20.1\pm0.4$ &           $0.88\pm0.01$ &            $0.262\pm0.003$ \\
0.85  &            $0.071\pm0.006$ &                                 0.35 &                                 7.5 &          $17.6\pm0.3$ &           $0.72\pm0.02$ &            $0.236\pm0.003$ \\
0.95  &            $0.119\pm0.008$ &                                 0.34 &                                 4.8 &          $14.8\pm0.3$ &           $0.53\pm0.02$ &            $0.139\pm0.002$ \\
\hline
\end{tabular}
    \caption{Spectral parameters of all ten phase-bins of \obj{} in phase-resolved analysis in three observations. The modelling details are described in sect.~\ref{results-ph-res}. The position of zero-phase is arbitrary. }
    \label{tab:ph-res}
\end{table*}


\bsp	
\label{lastpage}
\end{document}